\documentclass[aps,twocolumn,showpacs]{revtex4}
\usepackage[utf8x]{inputenc}
\usepackage[english]{babel}
\usepackage{hyperref}
\usepackage{blindtext}
\usepackage{bm}
\usepackage{amsmath,amsthm}
\usepackage{amssymb}
\usepackage{mathrsfs,dsfont,nicefrac}

\usepackage{epsfig}
\usepackage{graphicx}
\usepackage{subfig}

\usepackage{hyperref}

\providecommand{\braket}[2]{\ensuremath{\langle #1| #2 \rangle}}
\providecommand{\ketbra}[2]{\ensuremath{| #1 \rangle\! \langle #2 |}}
\providecommand{\bra}[1]{\ensuremath{\langle #1|}}
\providecommand{\ket}[1]{\ensuremath{|#1\rangle}}
\providecommand{\abs}[1]{\ensuremath{\lvert#1\rvert}}
\providecommand{\norm}[1]{\ensuremath{\lVert#1\rVert}}

\begin{document}
\title{Environment-Assisted Error Correction of Single-Qubit Phase
  Damping}

\author{Benjamin Trendelkamp-Schroer}
\author{Julius Helm}
\author{Walter T.~Strunz}

\affiliation{Institut f\"{u}r Theoretische Physik, Technische
  Universit\"{a}t Dresden, 01062 Dresden, Germany}

\date{\today}

\begin{abstract}
  Open quantum system dynamics of random unitary type may in principle
  be fully undone. Closely following the scheme of
  environment-assisted error correction proposed by Gregoratti and
  Werner [M.~Gregoratti and R.~F.~Werner, {\it J. Mod. Opt.} {\bf 50}
  (6), 915--933 (2003)], we explicitly carry out all steps needed to
  invert a phase-damping error on a single qubit. Furthermore, we
  extend the scheme to a mixed-state environment. Surprisingly, we
  find cases for which the uncorrected state is closer to the desired
  state than any of the corrected ones.
  
\end{abstract}


\maketitle
\section{Introduction}
Coherent superpositions of quantum states lay the foundation for
genuinely non-classical phenomena such as entanglement or interference
of massive particles. With growing system size these superpositions
become however increasingly fragile, thereby impeding future
realizations of quantum technology. The interaction of the system of
interest with its surroundings leads to a loss of the interference
potential---a process usually denoted with decoherence
\cite{Joos2002}. Phase damping (a.k.a. dephasing) denotes the case of
pure decoherence. Here, the populations remain unaffected; rather,
only the coherences itself (i.e., the off-diagonal elements of the
density matrix) are subject to decay. This discrimination of
decoherence against pure decoherence may for example be adequate in
situations where the characteristic time describing polarization decay
is much smaller than the typical dissipation time
\cite{Braun:2001p2029}.

Arguably, one may identify two main mechanisms responsible for
decoherence. The first and most prominent approach rests on the
language of open quantum systems \cite{Zurek2003,
  Schlosshauer2007}. Here, the system of interest is seen as being
part of a larger (often infinite) quantum system, where also the
environmental degrees of freedom are incorporated. The decoherence is
then a direct consequence of growing correlations between system and
environment. It is worth noting that these correlations are not
necessarily of quantum nature: the system may in fact decohere
completely without being entangled to its surroundings at all
\cite{EisertPlenio2002, PerniceStrunz2011}. As a second relevant
source of decoherence one may consider stochastic fluctuations of
ambient fields (or ``random external fields''
\cite{Nielsen_Chuang_2007, AlickiLendi1987}). In fact, fluctuating
fields were successfully identified as the main source of decoherence
in ion trap quantum computers, where fluctuations are present both in
the trapping field and in the lasers addressing the individual ions
\cite{Monz:2010p3941}. This ensemble-based approach may be described
in terms of a stochastic Schr\"{o}dinger equation, where time
evolution is unitary, yet stochastic. Usually, dynamics of this type
is denoted random unitary (RU).

In this context it is sometimes argued that the latter mechanism
should be merely seen as a fake decoherence process. The argument is
partly based on the idea that, due to the unitary character of time
evolution for individual members of the ensemble, the dynamics is in
principle reversible \cite{Zurek2003, Schlosshauer2007} (the spin echo
technique, for example, is based on this idea). However, it is known
that genuine open quantum system dynamics may be reversible, too,
provided the reduced dynamics is RU and thus indistinguishable from
the dynamics caused by fluctuating fields
\cite{gregoratti2003quantum}. Here, the actual correction procedure is
conditional on classical information obtained from a measurement
performed on the system's quantum environment. It thus represents an
instance of environment-assisted error correction.  In principle, any
phase damping of a single qubit or qutrit may be corrected this way,
since the dynamics is always of RU type \cite{LandauStreater1993,
  BuscemiDAriano2005}.

The present article provides a thorough assessment of all the steps
involved in a complete reversal of a given phase-damping dynamics on a
single qubit. In Sec.~\ref{sec:envir-assist-error} we discuss the
basic tools needed for the environment-assisted error correction of RU
dynamics. The scheme is then applied in Sec.~\ref{sec:expl-impl-corr}
to the example of single-qubit phase damping. In addition to the
original scheme, where the environment is assumed to be in a pure
state initially, we study a possible extension of the correction
scheme in case of a mixed-state environment
(Sec.~\ref{sec:mixed-environments}).

\section{Environment-assisted error correction}
\label{sec:envir-assist-error}
In the theory of quantum error correction one usually assumes that
there is a certain (low) probability for an error acting locally on
qubits or gates. Furthermore it is assumed that one needs to account
for different kinds of errors represented by a certain set of error
operations. The probabilities for different errors to occur are
independent of each other and given a priori
\cite{Gottesman:1997zz}. The correction mechanisms are designed such
that they can detect and correct all errors from the set of possible
error operations. This requires an encoding of the logical qubits used
for the actual computation into a larger number of physical qubits as
can be nicely seen in the Shor code picture \cite{Shor1995}. In
addition, the realization of a unitary gate in the algorithm requires
a translation into a set of unitaries acting on the physical
qubits. The whole process is described by the beautiful theory of
fault tolerant quantum computation
\cite{Nielsen_Chuang_2007}. Unfortunately as pointed out in,
e.g.\cite{Gottesman:1997zz}, the scaling of the number of physical
qubits is, though only polylogarithmic in the asymptotic limit, too
severe even for a very small number of computational qubits to permit
an experimental realization by current means. Sometimes, being able to
perform measurements on the environment of the qubits allows to
recover quantum states. As an example, in \cite{Alber2001} a
continuous monitoring of the decay processes provides the information
for a conditional recovery operation. 

\emph{Environment-assisted error correction} as proposed in
\cite{gregoratti2003quantum} is a different approach to quantum error
correction. Here, the correction is based on classical information
obtained from a measurement on the system's environment. The scheme
allows for a complete correction of RU dynamics, provided the
(quantum) environment is initially in a pure state. We closely follow
the ideas presented in \cite{gregoratti2003quantum}; we are, however,
interested in carrying out the procedure lined out in the
aforementioned work for a feasible realization of an open quantum
system. In the following we will present a scheme that allows the
explicit construction of an observable for such a measurement. Our
results will provide a correction scheme for a phase-damping
interaction between a qubit and an initially pure environment of
finite dimension.

\subsection{Quantum channels}
In the context of open quantum systems, a quantum channel describes
the time-evolution of a quantum system arising from the joint unitary
evolution of system plus environment. Under the usual assumption of
vanishing initial correlations, system and environment are initially
described by a product state $\rho \otimes \rho_E$. The channel is
then obtained by tracing out the environmental degrees of freedom
\begin{equation}\label{eq:CP_induced_map}
  \Phi(\rho)=\mathrm{tr}_E\left(e^{-\mathrm{iH}t}\rho\otimes\rho_E e^{\mathrm{iH}t}\right).
\end{equation}

The total Hamiltonian $\mathrm{H}$ includes the interaction between
system and environment. According to \cite{Kraus1969,Kraus1970}, every
such channel has a nonunique decomposition
\begin{eqnarray}
  \label{eq:Kraus-representation}
  \rho' := 
  \Phi(\rho) =
  \sum_{\alpha}\mathrm{K}_{\alpha} \rho \mathrm{K}^{\dagger}_{\alpha} =
  \sum_{\beta}\mathrm{L}_{\beta}\rho \mathrm{L}^{\dagger}_{\beta}, 
\end{eqnarray}
where two corresponding sets of so-called Kraus operators $K_\alpha,
L_\beta$ are related via a unitary matrix $\mathrm{V} =
\left(v_{\alpha \beta}\right)$ \cite{choi1975}, so that
\begin{equation}\label{eq:unitary_equivalence_Kraus_operators}
  \mathrm{K}_{\alpha}=\sum_{\beta}v_{\alpha\beta}\mathrm{L}_{\beta}.
\end{equation}

In case of RU dynamics it is possible to find a decomposition of the
channel into unitary operators,
\begin{eqnarray}
  \label{eq:random-unitary}
  \Phi_{\mathrm{RU}}(\rho) = 
  \sum_{\alpha} p_{\alpha} \mathrm{U}_{\alpha} \rho \mathrm{U}_{\alpha}^\dagger, 
\end{eqnarray}
where $p_{\alpha} > 0, \sum_{\alpha} p_{\alpha} = 1$. The Kraus
operators are thus unitary up to normalization, $K_{\alpha} =
\sqrt{p_{\alpha}} U_{\alpha}$. Obviously, a RU channel is unital, that
is, leaving the completely mixed state invariant,
$\Phi_{\mathrm{RU}}(\mathds 1) = \mathds 1$.

Phase damping (or dephasing) is the case of pure decoherence, where,
in a fixed basis $\{ \ket{n} \}$ (the phase-damping basis), no
population transfer occurs. According to this basis of ``preferred
states'', the projectors are constants of the motion. In such a case
the Hamiltonian describing the system-environment coupling may be
diagonalized with respect to the phase-damping basis:
$\mathrm{H}=\sum_{n}\ketbra{n}{n}\otimes \mathrm{h}_n$
\cite{GorinStrunz2004}. Here, the relative Hamiltonians $\mathrm{h}_n$
act on the environment, only.  Assuming the environment to start in a
pure state, $\ket{\psi_0}$, the phase-damping dynamics is fully
described in terms of the overlap of the relative states $\ket{\psi_n}
:= e^{-i \mathrm{h}_n t} \ket{\psi_0}$, that is
\begin{eqnarray}
  \label{eq:relative-states}
  \rho_{mn}' = \braket{\psi_n}{\psi_m} \; \rho_{mn}. 
\end{eqnarray}
Leaving the diagonal elements intact, phase-damping channels are
unital by definition. In their article \cite{LandauStreater1993}
Landau and Streater show that a phase-damping channel acting on a
single qubit or qutrit may always be decomposed into a RU
decomposition, Eq.~(\ref{eq:random-unitary}). In principle,
phase-damping errors on systems of dimension $2$ or $3$ may thus be
completely undone.

\subsection{The correction scheme}
If the initial state of the environment is pure,
$\rho_E=\ketbra{\psi_0}{\psi_0}$, a certain Kraus decomposition is
selected by choosing a basis $\{\ket{\chi_{\beta}}\}_{\beta}$ of
$\mathcal{H}_E$. Upon insertion into Eq.~(\ref{eq:CP_induced_map}),
this leads to
\begin{eqnarray}
  \label{eq:kraus-selection}
  \Phi(\rho) &=&
  \sum_{\beta} \bra{\chi_{\beta}} e^{-\mathrm{iH}t} \ket{\psi_0}\rho 
  \bra{\psi_0}e^{\mathrm{iH}t} \ket{\chi_{\beta}}  \nonumber \\
  &=& \sum_{\beta} L_{\beta} \rho L_{\beta}^\dagger, 
\end{eqnarray}
where we identify $\mathrm{L}_{\beta} =
\bra{\chi_{\beta}}e^{-\mathrm{iH}t}\ket{\psi_0}$.

In principle, this identification allows to select the dynamics where
only a single term of the operator sum (\ref{eq:Kraus-representation})
applies:
\begin{equation}
  \mathrm{L}_{\beta}\rho \mathrm{L}^{\dagger}_{\beta}=
  \mathrm{tr}_E\left(e^{-\mathrm{iH}t}\rho\otimes\rho_E 
    e^{\mathrm{iH}t} (\mathds{1}\otimes \mathrm{P}_{\beta} )\right), 
\end{equation}
with the projector $\mathrm{P}_{\beta} =
\ketbra{\chi_{\beta}}{\chi_{\beta}}$. The unitary equivalence between
different decompositions allows to write $\mathrm{K}_{\alpha} =
\bra{\mu_{\alpha}} e^{-\mathrm{iH}t} \ket{\psi_0}$ with
$\ket{\mu_{\alpha}} = \sum_{\beta} \mathrm{V}_{\alpha\beta}
\ket{\chi_{\beta}}$.

The principle of correction of a RU channel $\Phi_{RU}$ is now
straightforward: it relies on the identification of an appropriate
basis $\ket{\mu_{\alpha}}$ corresponding to a RU decomposition of the
channel. The projectors $\mathrm{Q}_{\alpha} =
\ketbra{\mu_{\alpha}}{\mu_{\alpha}}$ may then be used to single out
the subnormalized sub-state
\begin{eqnarray*}
  \rho'_{\alpha} = 
  p_{\alpha}\mathrm{U}_{\alpha} \rho \mathrm{U}^{\dagger}_{\alpha}
  =\mathrm{tr}_E\left(e^{-\mathrm{iH}t}\rho\otimes\rho_E
    e^{\mathrm{iH}t} (\mathds{1}\otimes
    \mathrm{Q}_{\alpha})\right), 
\end{eqnarray*}
which, upon normalization, is unitarily related to the initial state
$\rho$.

For a suitable observable $\mathrm{O}$ on $\mathcal{H}_E$ with
projectors on non-degenerate eigenspaces,
$\mathrm{O}=\sum_{\alpha}\lambda_{\alpha}\mathrm{Q}_{\alpha}$ with
$\mbox{tr}(\mathrm{Q}_{\alpha} \mathrm{Q}_{\beta} ) =
\delta_{\alpha\beta} \mathrm{Q}_{\alpha}$, a measurement outcome of
$\lambda_{\alpha}$ allows to unambiguously discriminate between
sub-states $\rho'_{\alpha}$. The perfect recovery of the initial state
is achieved by $\mathrm{U}^{\dagger}_{\alpha} \rho_{\alpha}
\mathrm{U}_{\alpha}$. Gregoratti and Werner show in
\cite{gregoratti2003quantum} that RU channels are the only ones thus
allowing for a perfect correction.

\section{Explicit implementation: Correction of single qubit phase damping}
\label{sec:expl-impl-corr}
In practice, the correction scheme faces several impediments. First of
all, no simple method is known how to decide whether a given channel
has a RU decomposition. Second, even if such a decomposition is
possible, the single unitaries $\mathrm{U}_{\alpha}$ have to be known
in detail. Once these hurdles are overcome, one has to find explicit
expressions for all $\mathrm{Q}_{\alpha}$. The realization of this
last step is discussed in the original formulation
\cite{gregoratti2003quantum}. Here, the unitary equivalence between
equivalent Kraus decompositions,
Eq.~(\ref{eq:unitary_equivalence_Kraus_operators}), plays a central
role.

Now we have all ingredients to construct a Hamiltonian generating a RU
channel on a qubit and to proceed with the construction of the
observable. For a phase-damping channel on a qubit there are only two
relative Hamiltonians $\mathrm{h}_1$, $\mathrm{h}_2$ leading to the
relative states $\ket{\psi_1(t)}=e^{-\mathrm{ih_1}t}\ket{\psi_0}$,
$\ket{\psi_2(t)}=e^{-\mathrm{ih_2}t}\ket{\psi_0}$ on the
environment. The action of the phase-damping channel has a very simple
form in terms of their overlap, $C(t):=\braket{\psi_2(t)}{\psi_1(t)}$,
\begin{equation}\label{eq:action_dephasing_channel_pure}
  \rho' =
  \left(\begin{array}{cc} 
      \rho_{11} & C \rho_{12} \\ 
      \bar{C} \rho_{21} & \rho_{22} 
    \end{array}\right). 
\end{equation}
We have suppressed the time dependence in the equation above and will
from now on only consider quantities for a fixed value of $t$. We have
already seen that using the basis $\{\ket{\chi_{\beta}}\}_{\beta}$ of
$\mathcal{H}_E$ to explicitly perform the trace allows us to identify
$\mathrm{L}_{\beta}=\bra{\chi_{\beta}}e^{-\mathrm{iH}t}\ket{\psi_0}$
so that we can write
\begin{eqnarray}
  \label{eq:kraus-decomp-relative}
  \mathrm{L}_{\beta}=
  \left(
    \begin{array}{cc}
      \braket{\chi_{\beta}}{\psi_1} & 0 \\ 
      0 & \braket{\chi_{\beta}}{\psi_2} 
    \end{array}
  \right).
\end{eqnarray}

\subsection{RU decomposition}
In order to obtain the RU decomposition we compute the dynamical or
Choi matrix of our phase-damping channel. For
$n=\mathrm{dim}\mathcal{H}_S$, $\Phi:\mathcal{H}_S \rightarrow
\mathcal{H}_S$ we can identify $\mathcal{H}_S\sim \mathds{C}^n$ and
the Choi matrix is defined as the $n^2 \times n^2$ matrix containing
the action of $\Phi$ on all $n\times n$ matrices
$\mathrm{E}_{ij}=\mathbf{e}_i^T \mathbf{e}_j$ that form a basis of
$\mathcal{M}(n\times n, \mathds{C})$,
\begin{equation}
  \label{eq:Choi_matrix}
  \left(
    \begin{array}{cc}
      \Phi\left(\begin{array}{cc} 1 & 0 \\ 0 & 0\end{array}\right) & \Phi\left(\begin{array}{cc} 0 & 1 \\ 0  & 0\end{array}\right) \\ \Phi\left(\begin{array}{cc} 0 & 0 \\ 1  & 0 \end{array}\right) & \Phi\left(\begin{array}{cc} 0 & 0 \\ 0  & 1\end{array}\right) 
    \end{array}
  \right)=
  \left(
    \begin{array}{cccc} 
      1 & 0 & 0 & C \\ 
      0 & 0 & 0 & 0 \\ 
      0 & 0 & 0 & 0 \\ 
      \bar{C} & 0 & 0 & 1
    \end{array}
  \right).
\end{equation}
Diagonalization of this $4 \times 4$ matrix yields the nonzero
eigenvalues
\begin{itemize}
\item $\lambda_1 = 1-|C|$: 
  $\mathbf{v}_{\lambda_1}=\frac{1}{\sqrt{2}}(-\frac{C}{|C|},0,0,1)^T$
\item $\lambda_2 = 1+|C|$:
  $\mathbf{v}_{\lambda_2}=\frac{1}{\sqrt{2}}(\frac{C}{|C|},0,0,1)^T$
\end{itemize}
with corresponding eigenvectors $\mathbf{v}_{\lambda_1},
\mathbf{v}_{\lambda_2}$. According to a central result in
\cite{choi1975} we can obtain Kraus operators by rearranging the
eigenvectors of the above Choi matrix into $2 \times 2$ matrices,
resulting in the decomposition
\begin{eqnarray}\label{eq:RU_decomposition}
  \rho' & = & 
  \frac{1-|C|}{2}
  \left(
    \begin{array}{cc}
      -\frac{C}{|C|} & 0 \nonumber\\
      0 & 1 
    \end{array}
  \right) \rho 
  \left(
    \begin{array}{cc}
      -\frac{C^{*}}{|C|} & 0 \\ 
      0 & 1 
    \end{array}
  \right) \\ 
  && + \frac{1+|C|}{2}
  \left(
    \begin{array}{cc}
      \frac{C}{|C|} & 0 \\ 
      0 & 1 
    \end{array}
  \right) 
  \rho 
  \left(
    \begin{array}{cc}
      \frac{C^{*}}{|C|} & 0 \\ 
      0 & 1 
    \end{array}
  \right) \\
  &=:& \sum\limits_{\alpha=1}^2 K_\alpha \rho K_\alpha^\dagger. \nonumber 
\end{eqnarray}
Note that the Kraus operators of the decomposition in
(\ref{eq:RU_decomposition}) are trivially unitary up to a scaling
factor, $K_{\alpha} K_{\alpha}^\dagger = K_{\alpha}^\dagger K_{\alpha}
= p_{\alpha} \mathds 1, \alpha = 1,2$. It is easily verified that this
RU decomposition recovers
Eq.~(\ref{eq:action_dephasing_channel_pure}).

\subsection{Finding the correction observable}
Consider now the situation where the actual environment is of
dimension $n$. As a consequence the set
$\{\mathrm{L}_{\beta}\}_{\beta}$ contains $n$ elements and according
to Eq.~(\ref{eq:unitary_equivalence_Kraus_operators}) we know that
there exists a unitary matrix $\mathrm{V}$ that relates
$\{\mathrm{L}_{\beta}\}_{\beta}$ to the RU decomposition
$\{\mathrm{K}_{\alpha}\}_{\alpha}$ in (\ref{eq:RU_decomposition}), if
the latter is extended by $n-2$ zero matrices
$\mathrm{K}_{\alpha}=\sqrt{p}_{\alpha}\mathrm{U}_{\alpha}$,
$\alpha=1,2$, $\mathrm{K}_{\alpha}=0_{(2)}$, $\alpha=3,\dots,n$. We
will now show in detail how it is possible to obtain $\mathrm{V}$ from
the two sets of Kraus operators $\{\mathrm{K}_{\alpha}\}_{\alpha}$ and
$\{\mathrm{L}_{\beta}\}_{\beta}$. Due to their diagonal character the
Kraus operators may be rewritten in terms of vectors $K_{\alpha} =
(K^{\alpha}_{11}, K^{\alpha}_{22})$. The unitary equivalence,
Eq.~(\ref{eq:unitary_equivalence_Kraus_operators}), thus translates
to the following linear system for the rows $\mathbf{v}_{\alpha}$ of
$\mathrm{V}$,
\begin{equation}\label{eq:general_linear_system}
  \left(\begin{array}{c} K^{\alpha}_{1} \\ K^{\alpha}_{2} 
    \end{array}\right) = \left(\begin{array}{cccc} L^1_{1} & L^2_{1} & \dots & L^n_{1} \\ L^1_{2} & L^2_{2} & \dots & L^n_{2} \end{array}\right)\left(\begin{array}{c} v_{\alpha 1} \\ v_{\alpha 2} \\ \vdots \\ v_{\alpha n}  \end{array}\right), 
\end{equation}
where in addition all double indices are suppressed. For $\alpha=1,2$
(\ref{eq:general_linear_system}) is an inhomogeneous system
$\mathrm{A}\mathbf{v}_{\alpha}=\mathbf{b}_{\alpha}$ with
inhomogeneities obtained from $\mathrm{K}_1$, $\mathrm{K}_2$ given by
\begin{eqnarray}
  \label{eq:b_1}
  \mathbf{b}_1 &=\sqrt{\frac{1-|C|}{2}} \left(\begin{array}{c} -\frac{C}{|C|} \\ 1\end{array}\right) \\
  \label{eq:b_2}
  \mathbf{b}_2 &=\sqrt{\frac{1+|C|}{2}} \left(\begin{array}{c} \frac{C}{|C|} \\ 1 \end{array}\right).
\end{eqnarray}
For $\alpha=3,\dots,n$ one has to fulfill the homogeneous system
$\mathrm{A}\mathbf{v}_{\alpha}=0$. In addition, unitarity of
$\mathrm{V}$ requires that
$\braket{\mathrm{v}_{\alpha}}{\mathrm{v}_{\beta}}=\delta_{\alpha
  \beta}$. In other words we need to find $n-2$ orthonormal vectors
$\mathbf{v}_{\alpha}\,\in\,\mathrm{ker}(\mathrm{A})$ and two
orthonormal vectors
$\mathbf{v}_{\alpha}\,\in\,\mathrm{ker}(\mathrm{A})^{\perp}$
satisfying the inhomogeneous equations. In the following we will show
that the singular value decomposition of $\mathrm{A}$ will provide all
of the required solutions $\mathbf{v}_{\alpha}$.

The $2\times n$ matrix $\mathrm{A}$ has singular value decomposition
$\mathrm{A}=\mathrm{U}\Sigma \mathrm{W}^{\dagger}$, with $\mathrm{U}$
a unitary $2\times 2$ matrix with columns spanning $\mathrm{im(A)}$,
$\Sigma$ a $2 \times n$ diagonal matrix containing the singular values
$\lambda_i$ of $\mathrm{A}$, and $\mathrm{W}$ a $n\times n$ unitary
matrix. If $r=\mathrm{rank(A)}$, the first $r$ columns of $\mathrm{W}$
form an ONB of $\mathrm{ker(M)}^{\perp}$ and the last $n-r$ columns
form an ONB of $\mathrm{ker(A)}$.  Using the completeness relation
$\sum_{\beta=1}^n \ketbra{\chi_{\beta}}{\chi_{\beta}}=\mathds{1}$ it
is straightforward to show that
$\mathrm{A}\mathrm{A}^{\dagger}=\left(\begin{array}{cc}1 & C \\
    \bar{C} &
    1 \end{array}\right)=\mathrm{U}(\Sigma\Sigma^{\dagger})\mathrm{U}^{\dagger}$.
Diagonalizing $\mathrm{A}\mathrm{A}^{\dagger}$ results in
$\mathrm{A}\mathrm{A}^{\dagger}=U\left(\begin{array}{cc}1-|C| & 0 \\ 0
    & 1+|C| \end{array}\right)U^{\dagger}$, with eigenvectors $
\mathbf{u}_1 = \frac{1}{\sqrt{2}}\left( \begin{array}{c}
    -\frac{C}{|C|} \\ 1 \end{array}\right)$, $\mathbf{u}_2 =
\frac{1}{\sqrt{2}}\left(\begin{array}{c} \frac{C}{|C|} \\
    1 \end{array}\right)$ forming the columns of $\mathrm{U}$. The
singular values of $\mathrm{A}$ can be read off as the square root of
the eigenvalues of $\mathrm{A}\mathrm{A}^{\dagger}$,
$\lambda_1=\sqrt{1-|C|}$ and $\lambda_2=\sqrt{1+|C|}$. Now let
$\mathbf{w}_i$ be the columns of $\mathrm{W}$. Using the singular
value decomposition of $\mathrm{A}$ it is now obvious that
$\mathrm{A}\mathbf{w}_1=\lambda_1 \mathbf{u}_1=\mathbf{b}_1$,
$\mathrm{A}\mathbf{w}_2=\lambda_2 \mathbf{u}_2=\mathbf{b}_2$ and
$\mathrm{A}\mathbf{w}_i=0$ for all $i=3,\dots,n$. This means that the
singular value decomposition of $\mathrm{A}$ provides the desired
solutions $\mathbf{v}_{\alpha}$ to the linear system
(\ref{eq:general_linear_system}) via the columns $\mathbf{w}_i$ of
the unitary matrix $\mathrm{W}$. Thus we can finally conclude that the
unitary matrix $\mathrm{V}$ relating the RU decomposition of $\Phi_D$
to the decomposition with respect to the basis
$\{\ket{\chi}_{\beta}\}_{\beta}$ is given by
\[\mathrm{V}=\mathrm{W}^T.\]
We can transform the basis $\{\ket{\chi}_{\beta}\}_{\beta}$ to the
desired measurement basis $\{\ket{\mu_{\alpha}}\}_{\alpha}$ via
\[\ket{\mu_{\alpha}}=\sum_{\beta}(\mathrm{W}^{\dagger})_{\alpha \beta}\ket{\chi_{\beta}}.\]
The projectors $\mathrm{Q}_{\alpha}$ for the observable $\mathrm{O}$
on $\mathcal{H}_E$ are given by
$\mathrm{Q}_{\alpha}=\ketbra{\mu_{\alpha}}{\mu_{\alpha}}$.

Now the general correction scheme consists of the following steps: i)
Performe a measurement on the environment using the observable
$\mathrm{O}=\sum_{\alpha=1}^n\lambda_{\alpha}\mathrm{Q}_{\alpha}$,
with $\lambda_{1}\neq \lambda_2$ resulting in a post measurement state
$\rho_{\alpha}=\mathrm{U}_{\alpha}\rho \mathrm{U}^{\dagger}_{\alpha}$
for all possible measurement outcomes $\lambda_{\alpha}$. ii) Apply
the unitary transformation
$\mathrm{U}^{\dagger}_{\alpha}\rho_{\alpha}\mathrm{U}_{\alpha}$
conditional on the outcome of the measurement of $\mathrm{O}$. The
probability to get $\lambda_{\alpha}$ is
$p(\lambda_{\alpha})=p_{\alpha}$, $\alpha=1,2$ and
$p(\lambda_{\alpha})=0$ otherwise. 

\begin{figure}[t]
  \centering 
  \includegraphics[width=0.5\textwidth]{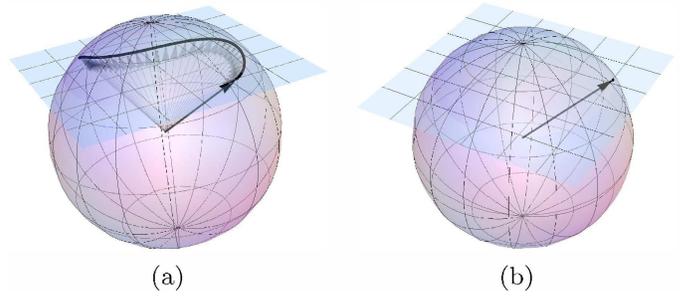}
  \caption{(Color Online) Phase-damping and correction for an
    initially pure environment. (a) The pure initial state $\rho$ of
    the system (black arrow) is mapped to a mixed state $\Phi_t
    (\rho)$ (thick, black line). The phase-damping interaction leaves
    invariant the $z$-value of the Bloch vector (the time-evolved
    states lie on a plane of constant $z$). (b) The corrected state
    $\tilde{\rho}=R_{\alpha}(\rho_{\alpha})$ after a measurement of
    $\mathrm{O}$ with outcome $\lambda_{\alpha}$ and application of
    the appropriate correcting channel $R_{\alpha}$. The corrected
    state $\tilde{\rho}$ is identical to the initial state $\rho$ of
    (a) for all of the above channels $\Phi_t$.}
  \label{fig:figure1}
\end{figure}

The following sequence of figures intends to provide further insight
into the mechanism of the correction scheme. For simplicity we
consider a two dimensional environment so that we can represent state
vectors of system and environment in the Bloch-sphere
picture. Fig.~\ref{fig:figure1} (a) shows a sequence of phase-damping
channels $\Phi_t$ acting on the pure initial system state $\rho$. The
adjacent Fig.~\ref{fig:figure1} (b) displays the corrected state for
each of the states $\Phi_t(\rho)$ after a measurement of the
correction observable $O$ (which in general will be different for
different channels $\Phi_t$) and the application of the correction
procedure $R_{\alpha}$ (which depends on the measurement outcome
$\alpha$ as well as the channel $\Phi_t$). The correction procedure
recovers the initial state $\rho$ for each channel $\Phi_t$. One can
think of the combined action of measurement and correction procedure
as the process that mediates between Fig.~\ref{fig:figure1} (a) and
(b).

\begin{figure}[h]
  \centering
  \includegraphics[width=0.48\textwidth]{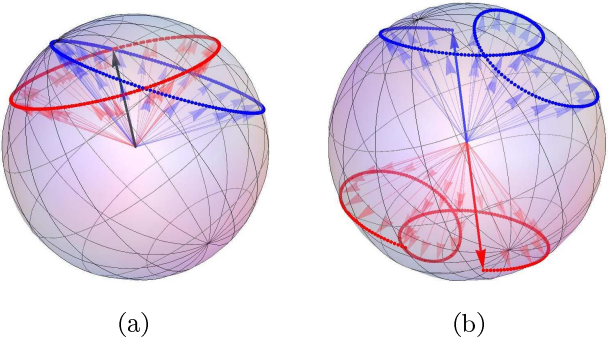}
  \caption{(Color online) (a) Unitary evolution of the relative states
    $\ket{\psi_1(t)}$ (red) and $\ket{\psi_2(t)}$ (blue). The initial
    state $\ket{\psi_0}$ is indicated by the black arrow lying on the
    intersection of the cones. (b) Correction observables
    $\mathrm{E}_1$ (red) and $\mathrm{E}_2$ (blue) for different
    channels $\Phi_t$. The two observables span orthogonal subspaces
    of $\mathcal{H}_E$ (the corresponding Bloch vectors point to
    opposite points).}
  \label{fig:figure2}      
\end{figure}

The phase-damping channel $\Phi_t$ is completely determined by the two
relative states $\ket{\psi_1(t)}$, $\ket{\psi_2(t)}$. Their unitary
time evolution is displayed in Fig.~\ref{fig:figure2} (a). The states
$\ket{\mu}_{\alpha}$ spanning the eigenspaces of the correction
observable $\mathrm{O}$ do also depend on the time parameter $t$
labeling the different phase damping channels $\Phi_t$. Their
time-evolution is displayed in Fig.~\ref{fig:figure2} (b).

\section{Mixed environments}
\label{sec:mixed-environments}
Until now it was assumed that the environment is initially in the pure
state $\ket{\psi_0}$. In this section we want to investigate the case
of a mixed-state environment. We apply modified versions of the
correction scheme for pure environments and explicitly show that none
of these protocols leads to a satisfying correction of the
phase-damping channel. For a mixed environment there is no unique
correspondence between the projection $\mathrm{Q}_{\alpha}$ on the
environment Hilbert space and a single Kraus operator
$\mathrm{K}_{\alpha}$ as in the case of a pure initial environment. We
assume that our environment is two-dimensional and that its initial
state is given by the mixture
$\rho_E=w\ketbra{\psi_0}{\psi_0}+(1-w)\ketbra{\psi^{\perp}_0}{\psi^{\perp}_0}$.
For concreteness think of a low but not zero temperature environment
such that $w$ is less but close to $1$. The coupling is realized via
the following Hamiltonian, $\mathrm{H}=k \sigma_z^{S}\otimes
\sigma_z^{E}+\mathds{1}\otimes \mathbf{\Gamma}\mathbf{\sigma}$,
$\sigma=\sum_{i=1}^3 \sigma_i \mathbf{e}_i$ the vector of Pauli-spin
matrices and $\mathbf{\Gamma} \in \mathds{R}^3$. The Hamiltonian can
be written as
$\sum_{i=1,2}\ketbra{s_z^{i}}{s_z^{i}}\otimes\mathrm{h}_i$, where
$\ket{s_z^{i}}$ are the eigenstates of $\sigma_z^S$ and
$\mathrm{h}_1=k\sigma_z^{E}+\mathbf{\Gamma}\sigma^E$,
$\mathrm{h}_2=-k\sigma_z^{E}+\mathbf{\Gamma}\sigma^E$. The channel
$\Phi$ generated by $\mathrm{H}$ according to
Eq.~(\ref{eq:CP_induced_map}) preserves the diagonal elements of
$\rho$ in the $\ket{s_z^{i}}$ basis. For $\rho_E$ mixed the
phase-damping interaction generates two additional relative states
$\ket{\psi^{\perp}_1}=e^{-\mathrm{ih_1}t}\ket{\psi^{\perp}_0}$ and
$\ket{\psi^{\perp}_2}=e^{-\mathrm{ih_2}t}\ket{\psi^{\perp}_0}$ in
$\mathcal{H}_E$. The action of $\Phi$ is similar to the pure case,
Eq.~(\ref{eq:action_dephasing_channel_pure}), but $C$ is replaced by
$wC+(1-w)C^{\perp}$. $C^{\perp}$ is again given by an overlap of
relative states, $C^{\perp}=\braket{\psi^{\perp}_2}{\psi^{\perp}_1}$.

\subsection{Correction schemes for mixed environments}
\begin{figure}[htb]
\centering
\includegraphics{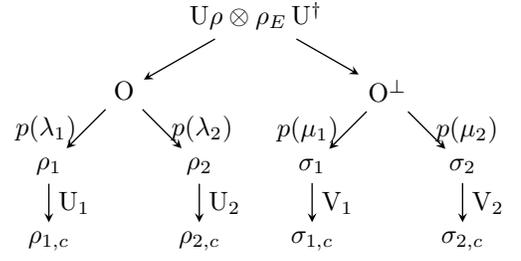}
\caption{Schematic picture of the possible correction schemes for a
  mixed environment (see text for details). }
\label{fig:figure3}
\end{figure}

In order to perform a correction we treat the channel as if it was
generated by the pure initial environment state $\ket{\psi_0}$ (or
$\ket{\psi^{\perp}_0}$). For $\ket{\psi_0}$ we calculate the
corresponding projectors $\mathrm{Q}_1$, $\mathrm{Q}_2$, the
observable $\mathrm{O}=\sum_{i=1,2}\lambda_i \mathrm{Q}_i$,
$\lambda_1\neq\lambda_2$, for a measurement on the environment,
probabilities $p (\lambda_1)$, $p (\lambda_2)$ of the RU decomposition
and unitary operators $\mathrm{U}_1$, $\mathrm{U}_2$ for the
correction operation. We will call the states resulting from this
correction procedure $\rho_{\alpha,c}$. Similarly for
$\ket{\psi^{\perp}_0}$ we calculate $\mathrm{P}_1$, $\mathrm{P}_2$,
$\mathrm{O}^{\perp}=\sum_{i=1,2}\mu_i \mathrm{P}_i$, $\mu_1 \neq
\mu_2$, $r_1$, $r_2$ and $\mathrm{V}_1$, $\mathrm{V}_2$ and call the
corrected states $\sigma_{\alpha,c}$ (the possible correction pathways
for a mixed state are displayed in Fig.~\ref{fig:figure3}). A
measurement of $\mathrm{O}$ ($\mathrm{O}^{\perp}$) and a successive
outcome dependent correction $\mathrm{U}_i$ ($\mathrm{V}_i$) will not
perfectly correct $\Phi$ since we have neglected the fact that
$\rho_E$ is a mixture.

Several correction protocols can be considered. First, the corrector
might simply ignore the fact that there is a small admixture of
$\ket{\psi^{\perp}_0}$. In this case, the state after correction is
given by $p (\lambda_1) \rho_{1,c} + p (\lambda_2) \rho_{2,c} =:
\rho_c$. Second, as a matter of keeping an open mind the corrector
might want to mirror the mixed-state probabilities for the two states
in his choice of correction scheme. In other words, he performs the
measurement on $\mathcal{H}_E$ and the successive correction with a
probability of $w$ as if $\rho_E$ was the pure state $\ket{\psi_0}$
and with a probability of $1-w$ as if $\rho_E$ was
$\ket{\psi^{\perp}_0}$. We call the state resulting from an ensemble
average over states corrected in this manner $\tilde \rho_c = w \rho_c
+ (1-w) \sigma_c$.

\subsection{The error of the correction}
\begin{figure}[htb]
  \centering
  \includegraphics[width=0.5\textwidth]{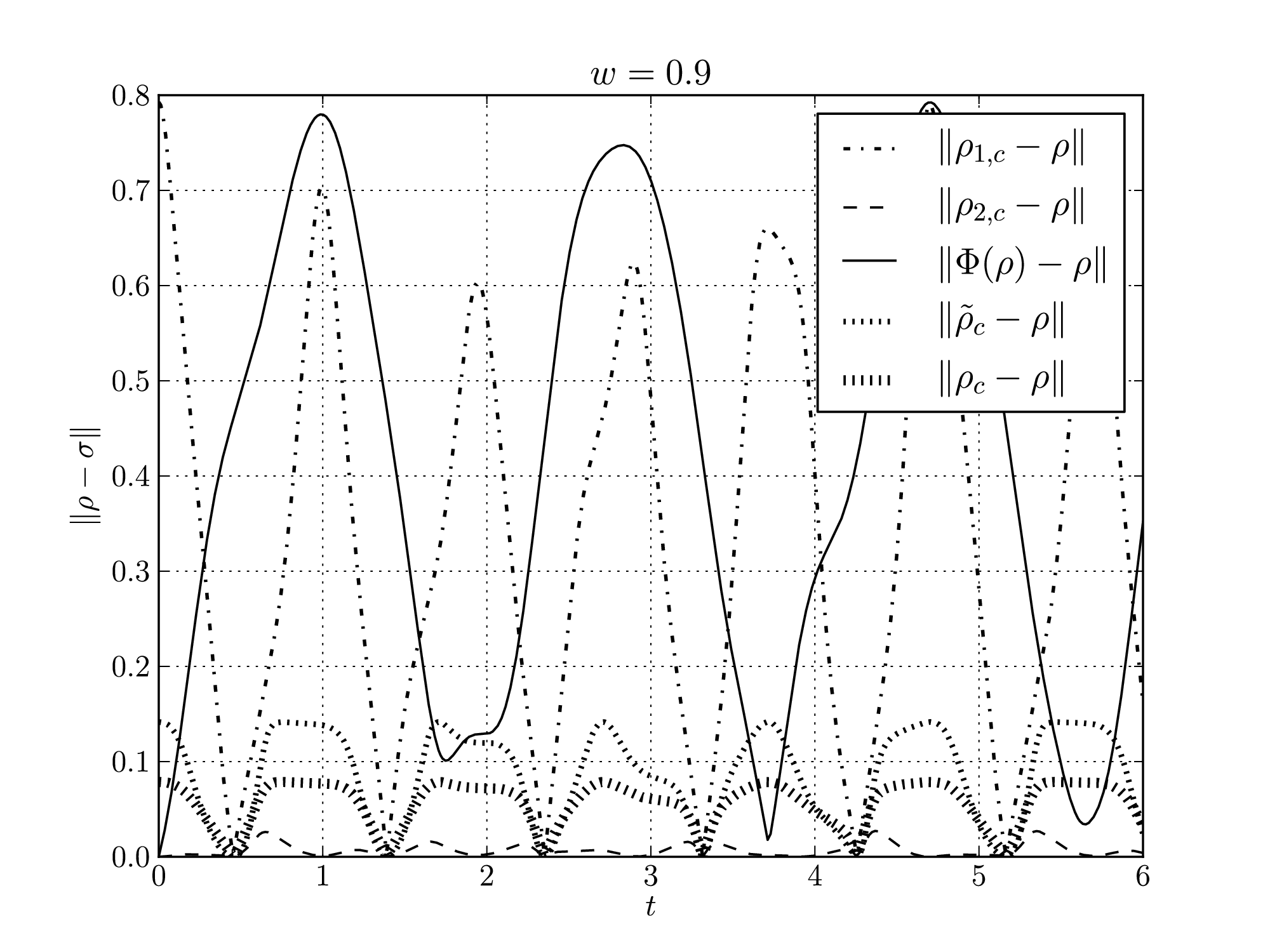}
  \caption{Trace distance as a measure for the quality of the
    corrected state. Different channels $\Phi_t$ arising from
    $\mathrm{H}$ and an initial state $\rho_E$ containing a small
    admixture of $\ket{\psi_0^{\perp}}$ ($w=0.9$) are corrected using
    the correction scheme for pure states (details see text). Clearly,
    there are cases where the uncorrected state $\Phi_t(\rho)$ (solid
    line) is closer to the desired state than the results from any of
    the correction schemes. }
\label{fig:figure4}
\end{figure}

In order to quantify the quality of the correction we measure the
distance between the corrected states $\rho_{\alpha, c}$,
$\sigma_{\alpha, c}$, $\rho_c$, and $\tilde \rho_c$ and the initial
state $\rho$, and compare it to the distance between the uncorrected
state $\Phi(\rho)$ and the initial state. The trace norm
\cite{Nielsen_Chuang_2007},
$\norm{\rho}=\frac{1}{2}\mbox{tr}\sqrt{\rho^{\dagger}\rho}$, is a
suitable distance measure in the set of states that will later allow
us to derive strict bounds analytically. In Fig.~\ref{fig:figure4} we
show results of the correction protocol for a small admixture of
$\ket{\psi^{\perp}_0}$ to $\ket{\psi_0}$ in $\rho_E$
($w=0.9$). Clearly, applying any of the correction procedures does not
necessarily lead to an improvement.  Indeed, we can see that even for
a small admixture there are phase-damping channels $\Phi_t$
(corresponding to interaction times $t$) for which the uncorrected
state is closer to the desired state than any of the corrected ones.

We observe from Fig. that the correction scheme is likely to worsen
the state in cases where the influence of the channel is small, i.e.,
$\norm{\rho-\Phi(\rho)} \rightarrow 0$. In order to shed light on this
worst case phenomenon we derive explicit expressions for the trace
distances. Channels that do not significantly alter the input state
can be characterized by assuming $C=1-\mathrm{i}\varepsilon$,
$C^{\perp}=1+\mathrm{i}\varepsilon$ with
$\mathcal{O}(\varepsilon^2)\approx 0$, see
App.~\ref{app:approximation_small_C}. For such channels it is possible
to find analytical expressions for the trace distances. An explicit
derivation can also be found in
App.~\ref{app:trace_distance_analytical}. For $\rho_E$ mixed, $w\neq1$
they are given (up to terms of $\mathcal{O}(\varepsilon^2)$) by
$\norm{\rho-\Phi(\rho)}=2\abs{\rho_{12}}\abs{w-\frac{1}{2}}\varepsilon$,
$\norm{\rho-\rho_{1,c}}=2
\abs{\rho_{12}}\abs{1+\mathrm{i}\varepsilon}$,
$\norm{\rho-\rho_{2,c}}=0$, and $\norm{\rho-\rho_{c}}=2
\abs{\rho_{12}}\abs{1-w} \abs{1+i \varepsilon}$. We can see that
\[\norm{\rho-\rho_{1,c}}\geq \norm{\rho-\rho_c} \geq
\norm{\rho-\Phi(\rho)}\geq \norm{\rho-\rho_{2,c}}.\] Furthermore, we
find $\norm{\rho-\tilde \rho_{c}}=2 \abs{\rho_{12}}\abs{
  w(1-w)+\frac{1}{2} }$ and, hence, \[\norm{\rho-\tilde \rho_c} \geq
\norm{\rho-\Phi(\rho)}.\] This allows to conclude that for channels
which do not significantly change the initial state $\rho$ the
uncorrected state $\Phi(\rho)$ is a better approximation to the
initial state than the corrected states $\rho_{c}$, $\tilde \rho_c$.

The inequality also shows that an improvement can be obtained by
selecting only the states $\rho_{2,c}$ corresponding to the very rare
outcome $\lambda_2$. Note, however, that for the corrector it is a
priori impossible to judge which correction scheme works best.

\section{Conclusion}
A phase-damping error of a single qubit arising from the interaction
with a pure and mixed finite-dimensional quantum environment is
considered. For the pure environment, we apply the correction scheme
proposed in \cite{gregoratti2003quantum} using knowledge about the
processes governing the full system dynamics and the initial state of
the environment. The correction is based on the RU decomposition for
every such channel from a spectral decomposition of the channels Choi
matrix. We are able to explicitly construct the unitary matrix
relating a decomposition with respect to an arbitrary basis of the
environment to the RU decomposition. The dynamics of such an open
quantum system can effectively be described using an environment of
only two dimensions spanned by the relative states $\ket{\psi_1}$ and
$\ket{\psi_2}$. 

How to correct a phase-damping error arising from an environment
starting in a mixed state is a delicate task. First, a sensible
correction protocol needs to be defined. We propose two rather
straightforward schemes and study their ability to correct. We observe
that in general the correction procedure for pure environments is no
longer successful. Indeed, there are cases for which the uncorrected
state is closer to the desired state than any of the corrected ones.
The surprising conclusion is that even small admixtures to the initial
state of the environment renders the success of the correction
undetermined.

\begin{acknowledgments}
  We would like to thank Sven Kr\"{o}nke for stimulating
  discussions. J.~H. acknowledges support from the International Max
  Planck Research School at the MPIPKS, Dresden. Thank is also due to
  the Alexander von Humboldt foundation and the Centro Internacional
  de Ciencias in Cuernavace, Mexico, for hosting a Humboldt Kolleg on
  open quantum systems where part of this work was completed.
\end{acknowledgments}

\begin{appendix}
  \section{The approximation for $|C| \approx
    1$ \label{app:approximation_small_C}}
  From the above discussion we know that requiring the phase-damping
  channel to be close to the identity is equal to saying that
  $wC+(1-w)C^{\perp}\approx 1$. Expanding the $2\times2$ Hamiltonian
  $\mathrm{h}_k$ in terms of Pauli matrices
  $\mathrm{h}=\sum_{i=1}^3\gamma_i \sigma_i$ and using the closed
  expression for the unitary operator, $e^{-\mathrm{ih}t}=\cos \alpha
  t \mathds{1}-\mathrm{i}\sin \alpha t \mathbf{n}\sigma$,
  $\alpha=\sqrt{\sum_i \gamma_i^2}$,
  $\mathbf{n}=\frac{\mathbf{\gamma}}{\alpha}$ we find that
  $\mathrm{Re}(C)=\mathrm{Re}(C^{\perp})$ and
  $\mathrm{Im}(C)=-\mathrm{Im}(C^{\perp})$. This means that for such
  channels $\mathrm{Re}(C)+\mathrm{i}(2w-1)\mathrm{Im}(C)\approx 1$
  and thus $C=1-\mathrm{i}\varepsilon$,
  $C^{\perp}=1+\mathrm{i}\varepsilon$ with
  $\mathcal{O}(\varepsilon^2)\approx 0$. Utilizing the closed
  expression for $e^{-\mathrm{ih}_k t}$ we can furthermore show that
  there are times for which the above requirements on $C$, $C^{\perp}$
  are simultaneously fulfilled.

\section{Analytical expressions for the trace distance \label{app:trace_distance_analytical}}
In order to derive analytical expressions for the trace distances we
express the post measurement states for a measurement of $\mathrm{O}$,
$\rho_{\alpha}$, and of $\mathrm{O^{\perp}}$, $\sigma_{\alpha}$, in
terms of the following channels. We define RU channels:
$\Phi_{RU,\alpha}(\rho)=\mathrm{tr}_E\left(\mathrm{U}\rho\otimes
  \ketbra{\psi_0}{\psi_0}\mathrm{U}^{\dagger}\mathds{1}\otimes
  \mathrm{Q}_{\alpha}\right)=p_{\alpha}\mathrm{U}_{\alpha}\rho
\mathrm{U}^{\dagger}_{\alpha}$,
$\Psi_{RU,\alpha}(\rho)=\mathrm{tr}_E\left(\mathrm{U}\rho\otimes
  \ketbra{\psi^{\perp}_0}{\psi^{\perp}_0}\mathrm{U}^{\dagger}\mathds{1}\otimes
  \mathrm{P}_{\alpha}\right)=r_{\alpha}\mathrm{V}_{\alpha}\rho
\mathrm{V}^{\dagger}_{\alpha}$ and auxiliary channels
$\Phi_{err,\alpha}(\rho)=\mathrm{tr}_E\left(\mathrm{U}\rho\otimes
  \ketbra{\psi_0}{\psi_0}\mathrm{U}^{\dagger}\mathds{1}\otimes
  \mathrm{Q}_{\alpha}\right)$,
$\Psi_{err,\alpha}(\rho)=\mathrm{tr}_E\left(\mathrm{U}\rho\otimes
  \ketbra{\psi^{\perp}_0}{\psi^{\perp}_0}\mathrm{U}^{\dagger}\mathds{1}\otimes
  \mathrm{P}_{\alpha}\right)$ so that we can express
$\rho_{\alpha}=\frac{w\Phi_{RU,\alpha}(\rho)+(1-w)\Phi_{err,\alpha}(\rho)}{\mathrm{tr}_E\left(w\Phi_{RU,\alpha}(\rho)+(1-w)\Phi_{err,\alpha}(\rho)\right)}$
and similarly $\sigma_{\alpha}$ replacing $\Phi$ with $\Psi$ and
interchanging $w$ and $1-w$. One can now see that the error of the
correction
$\rho_{\alpha,c}=\mathrm{U}^{\dagger}_{\alpha}\rho_{\alpha}\mathrm{U}_{\alpha}$
comes from the term
$(1-w)\mathrm{U}^{\dagger}_{\alpha}\Phi_{err,\alpha}(\rho)\mathrm{U}_{\alpha}$
and that $\rho_{\alpha,c}$ contains the desired state $\rho$ as $w
p_{\alpha}\rho$. A similar statement is also true for
$\sigma_{\alpha,c}$. In the following we will only discuss the states
$\rho_{\alpha,c}$ but a similar argument applies to
$\sigma_{\alpha,c}$. Using
$\mathrm{tr}\left(\Phi_{RU,\alpha}(\rho)+\Phi_{err,\alpha}(\rho)\right)=1$
we can verify that
$\mathrm{tr}\left(\Phi_{err,\alpha}(\rho)\right)=1-p_{\alpha}$. Observing
that
$p(\lambda_i)=w\mathrm{tr}\Phi_{RU,\alpha}(\rho)+(1-w)\mathrm{tr}\Phi_{err,\alpha}(\rho)$
we can use the above result to write down an analytical expression for
the probabilities of the measurement outcomes $\lambda_{\alpha}$ of
$\mathrm{O}$, $p(\lambda_\alpha)=w p_{\alpha} +
(1-w)(1-p_{\alpha})$. In the case that $C=1-\mathrm{i}\varepsilon$ we
can approximate the probabilities from the RU decomposition as
$p_1\approx0$, $p_2\approx1$. Using the fact that for positive
matrices the trace norm equals the trace,
$\mathrm{tr}(A)=\mathrm{tr}\sqrt{A^{\dagger}A}=\norm{A}_{nuc}$ and
that for finite dimensional matrices the trace norm is equivalent to
the maximum norm $\norm{A}_{\infty}=\underset{i,j}{\max}\,|a_{ij}|$ we
can conclude that $\Phi_{RU,1}(\rho)=\mathcal{O}(\varepsilon^2)$ and
$\Phi_{err, 2}(\rho)=\mathcal{O}(\varepsilon^2)$. Furthermore we can
express the unitary matrices from the RU decomposition as
$\mathrm{U}_1= \left(\begin{array}{cc} \mathrm{i}\varepsilon-1 & 0 \\
    0 & 1 \end{array}\right)$ and $\mathrm{U}_2=
\left(\begin{array}{cc} 1- \mathrm{i}\varepsilon & 0 \\ 0 &
    1\end{array}\right)$.  This allows to express the corrected states
as $\rho_{1,c}=\mathrm{U}^{\dagger}_1\Phi_{err, 1}(\rho)\mathrm{U}_1$
and $\rho_{2,c}=\mathrm{U}^{\dagger}_2\Phi_{RU,
  2}(\rho)\mathrm{U}_2$. We can already conclude that $\rho_{2,c}=
\rho+\mathcal{O}(\varepsilon^2)$. In order to obtain an expression for
$\Phi_{err, 1}(\rho)$ we use that
$\Phi(\rho)=\sum_{\alpha}\left(w\Phi_{RU,\alpha}(\rho)+(1-w)\Phi_{err,
    \alpha}(\rho)\right)$ and $\norm{\Phi_{err, 2}(\rho)}=1-p_2=0$ so
that we can approximate $\Phi(\rho)=(1-w)\Phi_{err, 1}(\rho)+w
\Phi_{RU,2}(\rho)$ and finally express $\Phi_{err,1}$ in terms of
$\Phi$ and $\Phi_{RU, 2}$. The final expression for the corrected
state $\rho_{1,c}$ is then given by
$\rho_{1,c}=\left(\begin{array}{cc}\rho_{11} &
    -(1+2\mathrm{i}\varepsilon)\rho_{12} \\
    -(1-2\mathrm{i}\varepsilon)\rho_{21} & \rho_{22}\end{array}
\right)$. It is now straightforward to calculate
$\norm{\rho-\rho_{1,c}}=2
\abs{\rho_{12}}\abs{1+\mathrm{i}\varepsilon}$,
$\norm{\rho-\Phi(\rho)}=2\abs{\rho_{12}}\abs{w-\frac{1}{2}}\varepsilon$,
and $\norm{\rho-\rho_{2,c}}=0$. For the corrected state $\rho_c$ we
obtain $\norm{\rho-\rho_c} = 2 \abs{\rho_{12}} \abs{1-w} \abs{1 + i
  \varepsilon}$. Performing a similar calculation for
$\sigma_{\alpha,c}$ and $p(\mu_{\alpha})$ one can obtain the following
expression for the distance between initial state and the corrected
state $\tilde \rho_c$: $\norm{\rho-\tilde \rho_{c}}=2
\abs{\rho_{12}}\abs{ w(1-w)+\frac{1}{2}}$.

\end{appendix}

\bibliographystyle{plain}

\end{document}